\documentclass[%
 aip,
 amsmath,amssymb,
 preprint,%
]{revtex4-1}

\usepackage{graphicx}
\usepackage{dcolumn}
\usepackage{bm}
\usepackage[mathlines]{lineno}

\usepackage[utf8]{inputenc}
\usepackage[T1]{fontenc}
\usepackage{mathptmx}
\usepackage{etoolbox}

\usepackage{subcaption}
\usepackage{xcolor}
\usepackage{tikz}
\usepackage{ulem}

\makeatletter
\def\@email#1#2{%
 \endgroup
 \patchcmd{\titleblock@produce}
  {\frontmatter@RRAPformat}
  {\frontmatter@RRAPformat{\produce@RRAP{*#1\href{mailto:#2}{#2}}}\frontmatter@RRAPformat}
  {}{}
}%
\makeatother
\begin{document}

\preprint{Physics of Fluids}

\title[Classifying acoustic cavitation with machine learning]{Classifying acoustic cavitation with machine learning trained on multiple physical models\footnote{This is the authors' peer-reviewed accepted manuscript, published in the AIP journal Physics of Fluids, volume 37, issue 3, article 037116 (March 2025). Please cite this article as DOI:10.1063/5.0255579}}
\author{Trinidad Gatica}
\affiliation{School of Engineering, Pontificia Universidad Católica de Chile, Santiago, Chile.}

\author{Elwin van 't Wout}
\email{e.wout@uc.cl}
\affiliation{Institute for Mathematical and Computational Engineering, School of Engineering and Faculty of Mathematics, Pontificia Universidad Católica de Chile, Santiago, Chile.}

\author{Reza Haqshenas}
\affiliation{Department of Mechanical Engineering, University College London, London, United Kingdom.}

\date{\today}

\begin{abstract}
Acoustic cavitation threshold charts are used to map between acoustic parameters (mainly intensity and frequency) and different regimes of acoustic cavitation. The two main regimes are transient cavitation, where a bubble collapses, and stable cavitation, where a bubble undergoes periodic oscillations without collapse. The cavitation charts strongly depend on the physical model used to compute the bubble dynamics and the algorithm for classifying the cavitation threshold. The differences between modeling approaches become especially noticeable for resonant bubbles and when sonication parameters result in large-amplitude oscillations. This paper proposes a machine learning approach that integrates three physical models, i.e., the Rayleigh-Plesset, Keller-Miksis and Gilmore equations, and multiple cavitation classification techniques. Specifically, we classify the cavitation regimes based on the maximum radius, the acoustic Mach number, the kurtosis factor of acoustic emissions, and the Flynn criterion on the inertial and pressure functions. Four machine learning strategies were developed to predict the likelihood of the transient and stable cavitation, using equally weighted contributions from classification techniques. By solving the differential equations for bubble dynamics across a range of sonication and material parameters and applying cross-validation on held-out test data, our framework demonstrates high predictive accuracy for cavitation regimes. This physics-informed machine learning approach offers probabilistic insights into cavitation likelihood, combining diverse physical models and classification strategies, each contributing different levels of physical rigor and interpretability.
\end{abstract}

\maketitle

\section{Introduction}

Acoustic waves propagating through a liquid can induce the formation, growth, and oscillation of vapor- or gas-filled cavities (bubbles) within the liquid~\cite{Leighton_1994}. The dynamical behavior of these bubbles strongly depends on the parameters of the acoustic waves (i.e., pressure anplitude, frequency, sonication protocol), the properties of the bubble, and the characteristics of the surrounding liquid. The two main cavitation regimes are stable and transient cavitation. Stable cavitation is characterized by periodic oscillations of a bubble. In contrast, a bubble undergoing transient cavitation rapidly expands to its maximum size, followed by a violent collapse. Mapping between acoustic parameters and cavitation regimes is important because different bubble dynamics generate different physical effects in the surrounding medium, including localized shock waves, micro-streaming and jetting, shear stresses, and the production of free radicals~\cite{Leighton_1994}. These effects have broad applications in science and engineering, such as sonochemistry~\cite{bhangu2017,suslick1999}, sonocrystallization~\cite{haqshenas2016modelling,haqshenas2018modelling,lee2018,chow2003}, and imaging and therapeutic ultrasound~\cite{averkiou2020,kooiman2020,coussios2008applications,stride2019nucleation}. 

Theoretical models of bubble dynamics are employed to conduct mechanistic studies and develop novel applications of acoustic cavitation. These models allow us to study stable and transient cavitation as a function of acoustic parameters, physical properties of the medium, and boundary conditions. The first theoretical models were developed in the early 1900s with the pioneering works of Lord Rayleigh~\cite{rayleigh1917} and Plesset~\cite{plesset1949}, which resulted in the development of the classical Rayleigh-Plesset equation. This ordinary differential equation (ODE) assumes that a bubble remains spherical during oscillations, there is no mass transport to/from the bubble, and the surrounding liquid is unbounded and incompressible. However, these assumptions break down near a wall or when damping becomes important in bubble dynamics, as observed in inertial cavitation~\cite{johnsen2009numerical}. Several other models have been developed to address these shortcomings~\cite{Gilmore_1952, Flynn_1975, Keller_Miksis_1980}. These theoretical models are nonlinear ODEs and are solved numerically using time-varying stiff ODE solvers. 

The exact properties of the bubble nuclei are often unknown in practical situations. Therefore, cavitation models are solved for a range of initial values, and the dynamics are classified using acoustic cavitation thresholds. These thresholds are important to determine the onset and type of cavitation in various scenarios. Classical examples include the mechanical index to predict the onset of inertial cavitation~\cite{apfel1991gauging} and indicators based on the maximum bubble radius or acoustic Mach number of a bubble oscillation~\cite{Leighton_1994}. The mechanical index is frequently used in ultrasound imaging to guide the selection of imaging parameters. Nonetheless, it is considered to be a poor indicator in many cases, especially when a bubble undergoes large-amplitude oscillations~\cite{miller2007}. As an alternative, the dynamical threshold based on radius-time curves was developed by Flynn~\cite{Flynn_1975}, which is more accurate than other thresholds but computationally more elaborate.

The common approach for analyzing cavitation characteristics is to select an appropriate theoretical model for the targeted application and solve it numerically for all expected combinations of acoustical and material parameters. The practical challenges of this approach are as follows: i) selecting a theoretical model that explains the physics correctly requires expert knowledge; ii) accurately solving these models with numerical integration schemes can be computationally expensive, especially when parameter sweeps are performed; iii) multiple cavitation thresholds should be computed as there is no single algorithm to predict cavitation types reliably. To address these challenges, this study proposes a machine-learning approach that combines multiple theoretical models and cavitation thresholds to predict the cavitation type quickly. This method provides physics-informed predictions incorporating multiple cavitation models and thresholds.

Machine learning algorithms have recently received increased interest as a predictive tool in computational acoustics~\cite{bianco2019machine, cuomo2022scientific}, with physics-informed neural networks~\cite{raissi2019physics} being the most well-known for multi-physics simulations~\cite{zhang2022multi, zhang2023multi, qiu2025adaptive}. In the context of acoustic cavitation, convolutional neural networks have been used to analyze images of bubble dynamics from laboratory experiments~\cite{orlova2023machine, marsh2024predicting}. Machine learning techniques were also proposed to simulate multiscale dynamics by training DeepONets on data generated with the Rayleigh-Plesset equation as a continuum model and direct simulation for particle dynamics~\cite{lin2021seamless}. However, this approach was only tested at acoustic pressures lower than typically encountered in practical applications of acoustic cavitation~\cite{lin2021operator}. A machine learning algorithm was also used to estimate the maximum bubble radius in a stable cavitation regime, trained on data generated by solving the Keller-Miksis equation~\cite{wang2021machine}. In our work, we introduce a novel machine-learning methodology for predicting cavitation regimes for a wide range of acoustical and material parameters relevant to biomedical ultrasound and sonochemistry applications. Our machine learning predictions determine the likelihood of stable or transient cavitation based on four different cavitation thresholds, which are presented as cavitation likelihood charts. This method offers a fast and accurate tool for designing sonication protocols. 

This manuscript first presents the physical and mathematical formulations in Section~\ref{sec: Formulations}. Specifically, that section covers the differential equations for bubble dynamics, the classifiers for cavitation type, the generation of the training data, the design of the machine learning strategies, and the performance metrics for the predictions. Subsequently, the computational results are presented in Section~\ref{sec: Results}, which shows the performance of the machine learning designs and the predictions of cavitation type. The conclusion of the research will be summarized in Section~\ref{sec: Conclusions}.

\section{Formulation}
\label{sec: Formulations}

This section explains the technical formulations of the study, including the theoretical models of bubble dynamics in Section~\ref{sec: bubble oscillation}, the cavitation classifiers in Section~\ref{sec: classification}, and the machine learning strategies in Section~\ref{sec: machine learning}.

\subsection{Bubble dynamics}
\label{sec: bubble oscillation}

A gas-filled bubble within a fluid medium will oscillate when subjected to an acoustic pressure field. These oscillations may remain stable over time or may be sufficiently strong to cause a collapse of the bubble. The dynamics of such movements in a Newtonian fluid can be modeled with the Navier-Stokes equation. We consider the following assumptions that are commonly used to model the dynamics of a single bubble: i) an unbounded fluid surrounds the bubble; ii) oscillations remain spherically symmetric; and iii) the mass and heat transfers between the bubble and the fluid are ignored. Notice that the first assumption restricts our study to an isolated bubble, with no acoustic interactions with other bubbles or a boundary. These assumptions enable us to describe the bubble's oscillations by its time-varying radius only. Furthermore, we assume that the relevant physical quantities depend on the temperature only and remain constant in time. Specifically, the parameter $\rho$ denotes the fluid's density in $\mathrm{kg}\cdot \mathrm{m}^{-3}$, $\sigma$ the surface tension in $\mathrm{N} \cdot \mathrm{m}^{-1}$, $\mu$ the fluid's viscosity in $\mathrm{kg} \cdot \mathrm{m}^{-1} \cdot \mathrm{s}^{-1}$, $p_v$ the vapor pressure in $\mathrm{kg} \cdot \mathrm{m}^{-1} \cdot \mathrm{s}^{-2}$, and $c$ the speed of sound in $\mathrm{m} \cdot \mathrm{s}^{-1}$. The time-dependent variables are $R$, the bubble radius in $\mathrm{m}$,  $\dot{R}$, the velocity of the bubble's surface, and $\ddot{R}$, the acceleration of the oscillations. The variables of the system's equilibrium state include $R_0$, the initial bubble radius, and $P_0$, the atmospheric pressure. Additionally, there is a known incident field, denoted as $P_{\text{inc}}$ and measured in $\mathrm{kg} \cdot \mathrm{m}^{-1} \cdot  \mathrm{s}^{-2}$.

There are different equations for modeling bubble dynamics. Here, we consider three most popular formulations: the Rayleigh-Plesset, Keller-Miksis, and Gilmore equations. A common term in these models is the difference between the internal and external pressure applied to the bubble's surface, given by
\begin{equation}
    P(t) = \left(P_{0}+\frac{2 \sigma}{R_{0}}-p_{v}\right)\left(\frac{R_{0}}{R}\right)^{3 \kappa}+p_{v}-\frac{2 \sigma}{R}-\frac{4 \mu \dot{R}}{R}-P_{0}-P_{\text{inc}},
    \label{equation: internal and external pressure}
\end{equation}
where $\kappa$ is the polytropic index of the gas. The incident field is modeled as a sine wave, given by
\begin{equation}
    P_{\text{inc}}(t) = P_a \sin(\omega t),
    \label{equation: incident field pressure}
\end{equation}
where $P_a$ denotes the amplitude, $\omega$ the angular frequency, and $t$ the time.

\subsubsection{Rayleigh-Plesset}

The classical Rayleigh-Plesset equation is given by~\cite{Leighton_1994}
\begin{equation}
    R \ddot{R}+\frac{3 \dot{R}^{2}}{2}=\frac{P}{\rho}.
    \label{equation: rayleigh-plesset}
\end{equation}
In addition to the abovementioned assumptions, this model assumes irrotational flow and incompressible liquid. These assumptions fail in the cases of high-amplitude oscillations and violent collapses.

\subsubsection{Keller-Miksis}

The Keller-Miksis equation~\cite{Keller_Miksis_1980} incorporates higher-order nonlinear terms that account for the finite speed of sound and the compressibility of the liquid. It is given by
\begin{equation}
     R \ddot{R}\left(1-\frac{\dot{R}}{c}\right)+\frac{3}{2} \dot{R}^{2}\left(1-\frac{\dot{R}}{3 c}\right)
    =\left(1+\frac{\dot{R}}{c}\right) \frac{P}{\rho}+\frac{R}{\rho c} \frac{\mathrm{d} P}{\mathrm{~d} t}.
    \label{equation: Keller-Miksis}
\end{equation}

\subsubsection{Gilmore}

The Gilmore equation~\cite{Gilmore_1952} uses the Tait's equation of state~\cite{tait1888report}, incorporating a variable speed of sound and accounting for the compressibility of the liquid. The Tait's equation represents the relation between pressure and density in a compressible liquid. We approximate the exponential in the Tait's equation by a polynomial of order $n$ and use $n=7$, based on experimental findings repoirted in the literature~\cite{li1967equation}. The Gilmore equation models the speed of sound as $C(P, t) = \sqrt{c^2 + (n-1) H}$, where the time-varying enthalpy $H=H(P,t)$ depends on the pressure. The Gilmore equation reads
\begin{equation}
    R \ddot{R} \left(1-\frac{\dot{R}}{C}\right) +\frac{3}{2}\dot{R}^2\left(1-\frac{\dot{R}}{3 C}\right) = \left(1+\frac{\dot{R}}{C}\right) H +\frac{R \dot{H}}{C}\left(1-\frac{\dot{R}}{C}\right).
    \label{equation: Gilmore}
\end{equation}
This differential equation is more suitable at higher velocities of bubble dynamics, a situation typically encountered at high-intensity acoustic fields~\cite{Gilmore_1952}.

\subsection{Classifying cavitation type}
\label{sec: classification}

Although cavitation is commonly classified as stable or transient, there is no consensus on exactly how to distinguish between these regimes. Various thresholds are proposed to classify the dynamics of bubbles. In this study, we consider four different classifiers, each based on distinct physical assumptions. They include acoustic emissions (i.e., irradiated pressure), maximum bubble radius and velocity, and pressure and inertial functions of the bubble dynamics. Four classifiers will be included in the machine learning algorithms. This approach ensures a more reliable classification of stable and transient events, as the methodology does not depend on a single threshold and the physical limitations behind the specific model. Instead, our approach combines multiple models and thresholds that characterize bubble dynamics.

\subsubsection{Dynamical threshold}
\label{dynamical threshold classifier}

The three ODEs in Section~\ref{sec: bubble oscillation} can be reformulated as
\begin{equation}
    \ddot{R} = IF + PF,
    \label{equation: decomposition inertial and pressure function}
\end{equation}
where $IF$ and $PF$ are called the inertial and pressure functions, respectively. For the Rayleigh-Plesset equation, Eq. \eqref{equation: rayleigh-plesset}, these functions become
\begin{align}
    & IF = - \dfrac{3}{2} \dfrac{\dot{R}^2}{R} ~, \\
    &  PF = \dfrac{P}{\rho R} ~.
    \label{eq: IF and PF flynn's version for RP}
\end{align}
For the Keller-Miksis and Gilmore equations, these functions have a similar structure but are more complex - not provided here for brevity. In words, we assign the terms that include $\dot{R}^2$ to the inertial function, and the pressure function collects the remaining terms which depend on the pressure and enthalpy. The formulation in Eq.~\eqref{equation: decomposition inertial and pressure function} indicates that the bubble dynamics are governed by competing functions associated with the inertia of the surrounding fluid and the pressure difference at the bubble's boundary together with the surface tension of the bubble. For the purpose of this classifier, we define the critical radius of a bubble as the radius where the inertial function intersects the minimum of the pressure function. In scenarios where they do not intersect, we compare the values of the two functions at the instant where $PF$ attains its minimum. Specifically, if $IF \leq PF$ at the minimum of $PF$, the inertial forces dominate the dynamics~\cite{Flynn_1975}, and we classify the cavitation as transient. Conversely, if $IF > PF$, no bubble collapse is expected, and the cavitation is classified as stable. Then, we determine the transition radius of a bubble by the energy dissipation during a compression cycle within the cavity \cite{Flynn_1975_1}. The procedure to obtain the transition radius is as follows: i) calculate the ratio of the mechanical work between two time intervals: throughout a complete cavitation cycle and during the compression stage; ii) search for the cycle with the highest energy dissipation; and iii) the transition radius is the maximum bubble radius in this cycle. The dynamical threshold is defined as the maximum between the critical and transition radii. This procedure was first described by Flynn~\cite{Flynn_1975}. We compare this value with the maximum radius of the bubble during the entire duration. If the maximum radius exceeds the dynamic threshold, the cavitation is classified as transient; otherwise, it is considered to be stable.

\subsubsection{Acoustic emissions}

An oscillating bubble irradiates pressure waves, which can be estimated using the following equation~\cite{Vokurka_1985}
\begin{equation}
    \left(\frac{P_{\text{{irr}}}}{P_{\infty}}-1\right) \frac{P_{\infty}}{\rho}=\frac{R}{r}\left(\ddot{R} R+2 \dot{R}^2\right)-\left(\frac{R}{r}\right)^4 \frac{1}{2} \dot{R}^2 ,
    \label{equation: acoustic emissions classifier}
\end{equation}
where $P_{\text{{irr}}}$ is the irradiated pressure or acoustic emission measured at the radial distance $r$ from the center of the bubble and $P_{\infty}$ is the pressure in the liquid at infinity. After solving the bubble dynamic, modeled by the mentioned ODEs, the acoustic emission is calculated. To analyze these time series, statistical parameters such as the kurtosis and crest factor~\cite{oppenheim1999discrete} are computed. Acoustic cavitation is classified as transient if the acoustic emissions exceed the thresholds of $13$ for kurtosis and $7$ for the crest factor; otherwise, it is classified as stable. These thresholds allow us to detect high-energy transient components in acoustic emissions, which manifest transient cavitation~\cite{haqshenas2015multi}.

\subsubsection{Mach number}

Transient cavitation is more likely when a bubble oscillates fast, i.e., the bubble wall velocity is large. Hence, we calculate the velocity $\dot{R}$ from the solution of the differential equations and find its maximum during the entire simulation. If the maximum velocity is higher than the speed of sound in the liquid (i.e., the acoustic Mach number defined as $\dot{R}/c$ is larger than one), the cavitation is classified as transient. Otherwise, we check if the initial radius is smaller than the radius calculated from the natural frequency of oscillation~\cite{neppiras1980acoustic} or not. If this condition is satisfied, the cavitation is classified as transient; and otherwise it is considered as stable~\cite{Leighton_1994}.

\subsubsection{Maximum radius}

The fourth classifier distinguishes transient from stable cavitation by comparing the maximum bubble radius with a critical radius. Here, we use a critical radius that depends on the acoustical and physical parameters~\cite{Leighton_1994}, which differs from the critical radius defined in Section~\ref{dynamical threshold classifier}. The critical radius $R_\mathrm{crit}$ is defined by
\begin{equation}
    \left(R_{\mathrm{crit}}\right)^{3 \kappa -1}=\frac{3 \kappa}{2 \sigma}\left(P_{0}+\frac{2 \sigma}{R_{0}}-p_{v}\right) R_0^{3\kappa}
\label{equation: critical radius rayleigh-plesset, keller-miksis}
\end{equation}
for the Rayleigh-Plesset and Keller-Miksis differential equations and 
\begin{equation}
    \left(R_{\mathrm{crit}}\right)^{3 \kappa -1}=\frac{3 \kappa}{2 \sigma}\left(P_{0}+\frac{2 \sigma}{R_{0}}\right) R_0^{3\kappa}
    \label{equation: critical radius gilmore}
\end{equation}
for the Gilmore equation. 
The bubble's oscillation is classified as transient if the maximum bubble radius exceeds the critical radius (i.e., $R_{max}>R_{crit}$; otherwise, it is classified as stable cavitation.

\subsection{Machine learning}
\label{sec: machine learning}

The previous sections presented three differential equations that model bubble dynamics and four classifiers to identify the cavitation regimes based on the bubble's oscillations. Rather than selecting a specific differential equation and classifier, we design machine learning algorithms that consider all information from the twelve possible combinations of differential equations and classifiers.

\subsubsection{Supervised machine learning design}
\label{sec: supervised machine learning}

Supervised machine learning methods are used in this work. Supervised machine learning refers to a family of techniques that predict a label from features. The label in supervised machine learning encodes the variable of interest, which is, in our case, a boolean indicating stable or transient cavitation. The features in supervised machine learning are the selected physical and acoustical parameters that encode known predictors for each simulation. Specifically, we use the initial radius, acoustic pressure, temperature, frequency, density, viscosity, surface tension, speed of sound, and vapor pressure as features.

Supervised machine learning algorithms learn patterns from known examples of feature-label pairs during a training phase. We create a dataset with training examples by selecting a range of input parameters, solving the three differential equations from Section~\ref{sec: bubble oscillation}, and applying the four classifiers from Section~\ref{sec: classification}. Each classifier can be applied to each simulation result. As a result, there are twelve combinations to create a training example of cavitation type for the same set of material and acoustical input parameters. In other words, each set of features has twelve values for the label.

This study aims to investigate the feasibility of machine learning in classifying cavitation types. The twelve training techniques allow us to design four supervised machine learning approaches. The creation of four different machine learning designs improves our understanding of the effectiveness of such strategies in more detail. Furthermore, all these designs facilitate the application of any classification and regression algorithm. 

\paragraph{Ensemble design}
The threshold-based ensemble design uses twelve independent machine learning algorithms. Each algorithm uses a specific combination of differential equations and classifiers to create a binary label for cavitation type. The predictions generated by this ensemble of twelve algorithms can be combined into a single result in two ways. First, we can average the individual results into a proportion. Second, we can apply majority voting to obtain a binary outcome, where ties are classified as stable cavitation.

\paragraph{Multi-objective design}
The multi-objective design uses a machine learning algorithm that predicts multiple labels for the same feature set. Specifically, for a given set of feature values, it predicts twelve labels. Each label corresponds to a specific combination of differential equations and classifiers. Like the ensemble design, averaging or majority voting can reduce the twelve results to a single outcome.

\paragraph{Expansion design}
The expanded data design considers the twelve combinations as independent data items. Hence, the training data set has twelve repetitions of the same features but with labels that may differ depending on the specific differential equation and classifier. The prediction is a boolean value for cavitation type.

\paragraph{Likelihood design}
The likelihood design uses the same training data structure as the expanded data design. However, the binary labels of transient and stable caviations are converted to the real values $0.0$ and $1.0$, respectively. Then, a regression algorithm predicts a real number. This outcome can be interpreted as the likelihood of having stable cavitation according to the set of differential equations and classifiers considered in this study.

\subsubsection{Generating training data}

The generation of training data for the machine learning designs involves solving the three theoretical models for bubble dynamics explained in Section~\ref{sec: bubble oscillation}, namely the Rayleigh-Plesset, Keller-Miksis and Gilmore equations. These differential equations are numerically solved with the LSODA time integrator~\cite{hindmarsh1983odepack}. Each experiment simulates the bubble dynamics for a duration of 20 periods of the incident acoustic wave. The numerical integrator has a resolution of $100\,000$ time steps in each acoustic period, which is sufficient to achieve high-precision calculations. Since the second-order differential equations are solved as a system of two first-order differential equations, the time integrator's output is the bubble's radius and velocity at each time step. These outcomes are then used to obtain the labels for the training data by applying the classifiers from Section~\ref{sec: classification} to the numerical simulations.

The accuracy of machine learning approaches strongly depend on the variety of training data. Here, this means that we must generate training examples for a wide range of physical and acoustical settings to our interest. We choose to vary four of the most significant parameters and keep the rest fixed. Specifically, we use ranges of the initial radius, frequency, pressure amplitude, and temperature relevant for typical engineering applications. See Table~\ref{tab: ranges for initial parameters for ML} for the values. We take 10 samples for each parameter, uniformly distributed within the ranges. This results in $10\,000$ unique combinations for the training dataset. 

\begin{nolinenumbers}
\begin{table}
    \caption{Ranges for the input parameters in the training dataset.}
    \label{tab: ranges for initial parameters for ML}
    \begin{tabular}{l c l}
    \hline
        Initial radius & $1- 20$ & $\mathrm{\mu m}$\\
        Acoustic pressure & $0.2 - 3$ & $\mathrm{M Pa}$\\
        Frequency & $0.02 - 2$ & $\mathrm{MHz}$\\
        Temperature & $10 - 60$ & $^{\circ}\mathrm{C}$\\
        \hline
    \end{tabular}
\end{table}
\end{nolinenumbers}

The material parameters are chosen to resemble water and calculated through standard state equations from the literature. Specifically, the temperature determines the density~\cite{deAndrade_Haqshenas_Pahk_Saffari_2021}, viscosity~\cite{Kestin_Sokolov_Wakeham_1978}, surface tension~\cite{deAndrade_Haqshenas_Pahk_Saffari_2021}, and speed of sound~\cite{DelGrosso_1970} of the medium. Furthermore, we consider the vapor pressure of the bubble to be $3270$ $\mathrm{Pa}$, the atmospheric pressure is $100$ $\mathrm{kPa}$, and the adiabatic index is~$1.33$.

\subsubsection{Performance metrics}

The performance of machine learning to predict the desired outcomes can be analyzed with different metrics. We use the accuracy, mean absolute error (MAE), and root mean squared error (RMSE) as performance metrics. Notice that we also calculate performance metrics for the intermediate results in the ensemble and multi-objective designs before taking the ensemble average. 

Specifically, the accuracy of a binary predictor is defined by
\begin{equation}
    \text{accuracy} = \frac{\text{number of correct predictions}}{\text{total number of predictions}}.
    \label{equation:definition_of_accuracy}
\end{equation}
The MAE is defined by
\begin{equation}
    \text{MAE} = \frac{1}{n} \sum_{i=1}^{n} \left| y_i - \hat{y}_i \right|,
\end{equation}
where $y_i$ represents the actual value, $\hat{y}_i$ the predicted value, and $n$ the number of predictions. Similarly, the RMSE is defined by
\begin{equation}
    \text{RMSE} = \sqrt{\frac{1}{n} \sum_{i=1}^{n} (y_i - \hat{y}_i)^2}.
\end{equation}
Notice that our machine learning designs include binary outcomes representing stable and transient cavitation and real-valued outcomes representing the likelihood of stable cavitation. In the case of binary outcomes, the MAE and RMSE are calculated by encoding the boolean as zero or one. In the other case of real-valued outcomes, the accuracy is calculated by rounding towards zero or one.

\subsection{Methodology}

The workflow of the proposed methodology consists of two main phases: training and prediction. In the training phase, we select values for key physical parameters—initial radius, acoustic pressure, frequency, and temperature—and solve cavitation models based on the Rayleigh-Plesset, Keller-Miksis, and Gilmore equations. Each simulation result is processed to classify bubble dynamics as stable or transient cavitation using the four thresholds outlined in Section~\ref{sec: classification}. This process is repeated for all input parameter sets, generating a dataset of feature-label pairs, where the features correspond to the physical parameters and the label represents the cavitation regime. A machine-learning model is then trained by fitting optimal parameters to this dataset. In the prediction phase, the trained model is used to classify the cavitation regime for a specific input set. Given values for the four physical parameters, the machine learning algorithm predicts the most likely cavitation regime: stable or transient.  Figure~\ref{table: workflow} shows the workflow.

As an example, the random-forest algorithm consists of various decision trees~\cite{breiman2001random}. Each decision tree is trained on a subset of the dataset and creates branches that minimize the entropy in each leaf node. During the prediction phase, each decision tree follows the branches for the input values and provides a binary label, with the final outcome being the majority vote on the cavitation regime.

\begin{nolinenumbers}
\begin{figure}[ht]
    \begin{tikzpicture}
        \node[text width=5cm, align=center] at (0, 0) {training};
        \node[draw, rounded corners, text width=5cm, align=center](box11) at (0, -1) {select input parameters};
        \node[draw, rounded corners, text width=5cm, align=center](box12) at (0, -2) {carry out ODE simulations};
        \node[draw, rounded corners, text width=5cm, align=center](box13) at (0, -3) {apply cavitation thresholds};
        \node[draw, rounded corners, text width=5cm, align=center](box14) at (0, -4) {create tabulated dataset};
        \node[draw, rounded corners, text width=5cm, align=center](box15) at (0, -5) {fit model parameters};
        \node[text width=5cm, align=center] at (6, 0) {prediction};
        \node[draw, rounded corners, text width=5cm, align=center](box21) at (6, -1) {select input parameters};
        \node[draw, rounded corners, text width=5cm, align=center](box22) at (6, -2) {infer cavitation regime};
        \node[draw, rounded corners, text width=5cm, align=center](box23) at (6, -3) {evaluate performance};
        \draw[->, thick] (box11.south) -- (box12.north);
        \draw[->, thick] (box12.south) -- (box13.north);
        \draw[->, thick] (box13.south) -- (box14.north);
        \draw[->, thick] (box14.south) -- (box15.north);
        \draw[->, thick] (box21.south) -- (box22.north);
        \draw[->, thick] (box22.south) -- (box23.north);
    \end{tikzpicture}
    \caption{Workflow of the machine learning methodology.}
    \label{table: workflow}
\end{figure}
\end{nolinenumbers}

\subsection{Supervised machine learning algorithm}

The proposed machine-learning framework is flexible and not restricted to any specific supervised learning algorithm for classification or regression tasks. Although a comprehensive comparison of different algorithms is beyond the scope of this article, Table~\ref{table:performance_metrics_for_algorithms} presents performance metrics for an ensemble design using five common algorithms. Based on these results and our computational experience across various settings, we selected the random forest algorithm for the subsequent analyses in this manuscript. This choice is motivated by three key factors: (i) The random forest consistently ranked among the top-performing algorithms in the test set, with performance metrics either the highest or within a few percentage points of the best; (ii) The reasonable gap between training and test scores indicates that the model is not overfitting; (iii) Hyperparameter tuning of the number of predictors showed that using 15 decision trees effectively balances precision and robustness.

\begin{nolinenumbers}
\begin{table}[bt!]
    \caption{Performance metrics for machine learning algorithms on train and test sets.}
    \label{table:performance_metrics_for_algorithms}
    \begin{tabular}{l c c}
        \hline
        \textbf{Algorithm} & \textbf{MAE Train} & \textbf{MAE Test} \\
        \hline
        Linear regression & 0.2805 & 0.2841 \\
        Logistic regression & 0.3058 & 0.3082 \\
        Decision tree & 0.0000 & 0.0862 \\
        Gradient boost & 0.0730 & 0.0785 \\
        Random forest & 0.0032 & 0.0870 \\
        \hline
    \end{tabular}
\end{table}
\end{nolinenumbers}

\section{Results}
\label{sec: Results}

This section presents the computational results of our machine learning designs to predict stable and transient cavitation types.

\subsection{Computational settings}

The differential equations governing bubble dynamics (see Section~\ref{sec: bubble oscillation}) were non-dimensionalized to improve computational robustness. Specifically, the radius was non-dimensionalized by the equilibrium radius $R_0$, resulting in the non-dimensional radius $R/R_0$. The non-dimensional time variable was based on the period of the acoustic wave. The ODEs are solved with the LSODA time integrator available in Python's Scipy library~\cite{virtanen2020scipy}. Since each simulation takes a few minutes on a standard desktop computer, we generated the training dataset by parallelizing the simulations over 32~cores on a high-performance compute node. This parallelization significantly reduced the overall computation time, allowing us to complete the data generation phase within a few hours only.

We implemented the machine learning algorithms with Python's Scikit-learn library~\cite{scikit-learn}, known for its robust and efficient tools for machine learning model development and evaluation. We analyze the performance of supervised machine learning by applying cross-validation. Specifically, a $5$-fold approach was employed, where the dataset was randomly partitioned into five subsets of equal size. In each fold, one subset was reserved for testing, while the remaining four subsets were used for training. This process was repeated across all five folds, ensuring that each subset served as the test set exactly once. The performance metrics for each fold were averaged to calculate the overall performance metrics.

\subsection{Bubble cavitation}

Two examples of cavitation that are indicative of the stable and transient cavitation are presented in this section. Figure~\ref{figure: stable cavitation time series} shows the evolution of an $R_0=10~\mu$m bubble, during $10$ periods of sonication. The sonication parameters are: pressure amplitude=$0.3$~MPa, and frequency=1.2 MHz. At this relatively low acoustic pressure, the bubble oscillates smoothly with a maximum radius smaller than $1.3R_0$, which is classified as stable cavitation. For this scenario, the three different physical models predict similar bubble dynamics. 

\begin{nolinenumbers}
\begin{figure}[ht]
    \includegraphics{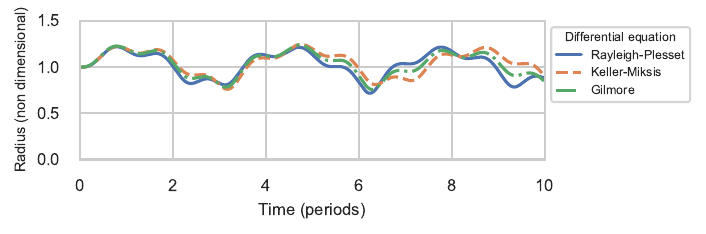}
    \caption{The oscillation of a bubble with $R_0=10~\mu$m, exposed to ultrasound waves with the frequency of 1.2~MHz and the amplitude of 0.3~MPa. This is an example of stable cavitation.}
    \label{figure: stable cavitation time series}
\end{figure}
\end{nolinenumbers}

Now, let us consider a scenario where $R_0=6~\mu$m, the sonication frequency and amplitude are $1.2$ MHz and $2$ MPa, respectively. Figure~\ref{figure: transient cavitation time series} shows that the bubble grows rapidly to a radius of around $4R_0$ within a cycle of excitation, followed by a strong collapse at the end of the second cycle. This type of bubble dynamics is a good example of transient cavitation. Furthermore, we can see that the predictions of the three models in the subsequent cycles are different, as the compressibility becomes important in this high-amplitude transient cavitation scenario.

\begin{nolinenumbers}
\begin{figure}[ht]
    \includegraphics{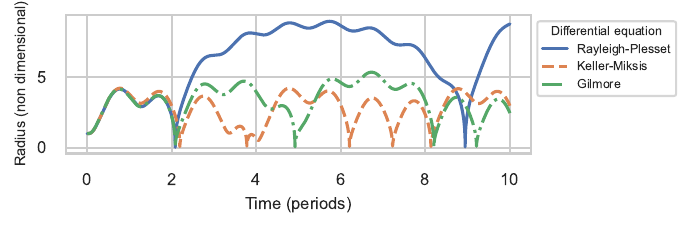}
    \caption{The oscillation of a bubble with $R_0 = 6~\mu$m, exposed to ultrasound waves with the frequency of 1.2~MHz and the amplitude of 2~MPa. This is an example of transient cavitation.}
    \label{figure: transient cavitation time series}
\end{figure}
\end{nolinenumbers}

\subsection{Training data}
\label{sec: training data}

We generate training data by solving the three models of the bubble dynamics presented in Section~\ref{sec: Formulations}. Then, four different classifiers are computed from these simulations. This procedure is repeated for $10\,000$ combinations of the four input parameters within the ranges given in Table~\ref{tab: ranges for initial parameters for ML}. These ten thousand numerical simulations, with twelve values for cavitation type each, form the training dataset for the supervised machine learning algorithms.

The labels in the training dataset have a distribution ratio of $60\%$ stable cavitation to $40\%$ transient cavitation. To explore this distribution in more detail, the acoustic cavitation data are grouped by the differential equation and classifier, as illustrated in Figures~\ref{fig:distribution_by_equation_of_data} and~\ref{fig:distribution_by_threshold_of_data}, respectively.

\begin{nolinenumbers}
\begin{figure}[ht]
    \includegraphics{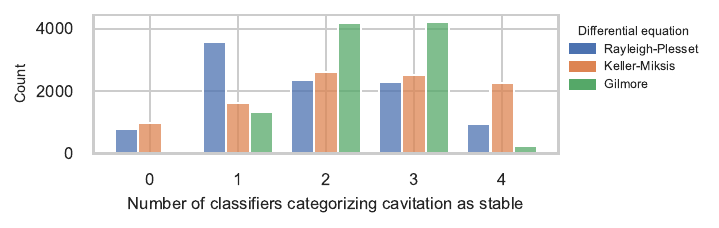}
    \caption{The distribution of acoustic cavitation data grouped by the differential equation. The horizontal axis represents the cumulative count of classifiers indicating stable cavitation events. The vertical axis counts the number of simulations, where the dataset has a total of $10\,000$ observations.}
    \label{fig:distribution_by_equation_of_data}
\end{figure}
\end{nolinenumbers}

The four classifiers in Section~\ref{sec: classification} all distinguish between stable and transient cavitation but through different modeling approaches and taking into account different physical assumptions. Therefore, the consistency between classifiers is not guaranteed. The horizontal axis in Figure~\ref{fig:distribution_by_equation_of_data} represents the consistency of classifiers. For example, the value zero refers to the cases where all classifiers judge the bubble cavitation as transient. Similarly, the value four represents the cases where all four classifiers consistently calssify cavitation as stable. These extreme cases have small counts (short bars), meaning that the complete agreement between the classifiers is rarely achieved. In fact, the largest proportion of cases have different classifiers giving different results. When looking at the differences between the bar groups, we notice that the Rayleigh-Plesset equation is more inclined to classify the same experiment as transient compared to the Keller-Miksis and Gilmore equations.

\begin{nolinenumbers}
\begin{figure}[ht]
    \includegraphics{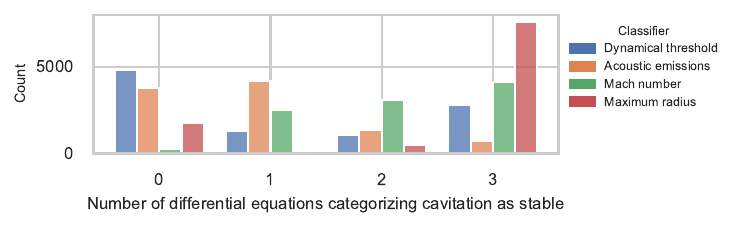}
    \caption{Distribution of acoustic cavitation data grouped by the classifier. The horizontal axis represents the cumulative count of differential equations indicating stable cavitation events. The vertical axis counts the number of simulations, where the dataset has a total of $10\,000$ observations.}
    \label{fig:distribution_by_threshold_of_data}
\end{figure}
\end{nolinenumbers}

Figure~\ref{fig:distribution_by_threshold_of_data} presents the label distribution of the training data grouped by the classifier. For example, the tallest bar at the right considers the maximum radius threshold as the classifier, for which, in more than $7000$ cases, from all three differential equations yield stable cavitation. Furthermore, the zero on the horizontal axis means that the simulations of the three differential equations feature transient cavitation. Differently, there is also a large proportion of situations where one differential equation yields another cavitation type than the other two differential equations. This inconsistency is due to the different modeling approaches of bubble dynamics and the physics of cavitation.

\subsection{Machine learning predictions}

We trained four different machine learning designs on the cavitation dataset, as explained in Section~\ref{sec: supervised machine learning}. Here, we present the performance results in predicting the cavitation type.

\subsubsection{Ensemble designs}

Let us first consider the ensemble and multi-objective designs, which both provide predictions for each of the twelve combinations of differential equations and classifiers. Figure~\ref{fig: accuracy for ensemble and multi-objetive} presents the accuracy of the machine learning predictions. The average accuracy is $91\%$ for both designs, and the prediction's accuracy differs slightly between the type of differential equation and classifier for cavitation. The errors in the machine learning's predictions are due to the training errors of the random forest algorithm but are also caused by the complexities in the data set. That is, the differential equations and classifiers for bubble cavitation also come with modeling errors.

\begin{nolinenumbers}
\begin{figure}[ht]
    \includegraphics{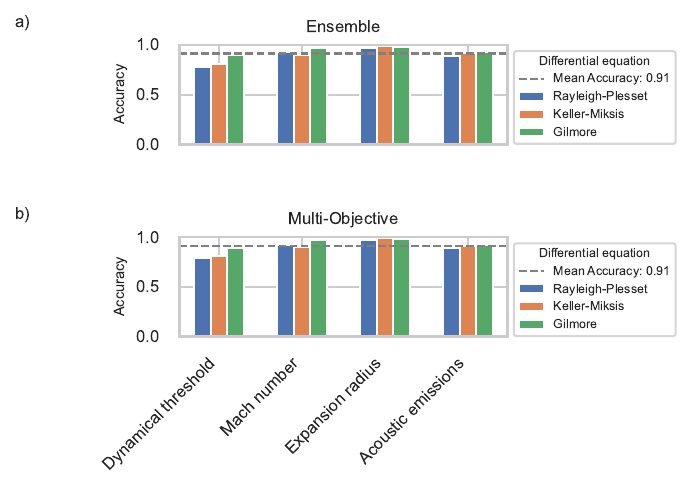}
    \caption{The accuracy score for ensemble and multi-objective designs for each of the twelve combinations between classifiers and differential equations.}
    \label{fig: accuracy for ensemble and multi-objetive}
\end{figure}
\end{nolinenumbers}

We observe that the dynamical threshold classifier has a lower accuracy than the other classifiers in both designs. This can be attributed to the classifier's more complex approach to distinguishing transient from stable cavitation from the bubble's oscillation profile. Hence, it is more challenging for machine learning to reproduce the cavitation behavior from the input parameters.

\subsubsection{Comparative performance}

Table~\ref{tab:final_performance_metrics} shows the performance metrics for all four machine learning designs, with the variants of the ensemble mean and majority voting for the first two. Generally speaking, the accuracy is reasonably high, and the errors sufficiently low. However, there are significant differences between the machine learning designs. For example, the ensemble and multi-objective models have a higher accuracy than the expansion and likelihood models. Remember that the first two designs train separate algorithms for specific combinations of differential equations and classifiers, while the latter two use a single machine learning algorithm for all combinations. This shows that it's easier for a machine learning algorithm to capture the behavior of a single differential equation and classifier, but it's more challenging to find patterns when all twelve combinations are included in the training set.

\begin{nolinenumbers}
\begin{table}[ht]
    \caption{Performance metrics for machine learning models on the test sets in five-fold cross-validation.}
    \label{tab:final_performance_metrics}
    \begin{tabular}{l c c c}
        \hline
        \textbf{Method} & \textbf{Accuracy} & \textbf{MAE} & \textbf{RMSE} \\
        \hline
        \text{Ensemble Mean} &  0.8230 & 0.0603 & 0.0894 \\
        \text{Ensemble Voting} & 0.8230 & 0.1770 & 0.4206 \\
        \text{Multi-objective Mean} & 0.8193 & 0.0610 & 0.0914 \\
        \text{Multi-objective Voting} & 0.8193 & 0.1807 & 0.4250 \\
        \text{Expansion} & 0.6116 & 0.3884 & 0.6232 \\
        \text{Likelihood} & 0.6251 & 0.0492 & 0.0689 \\
        \hline
    \end{tabular}
\end{table}
\end{nolinenumbers}

When considering the errors in the regression tasks of finding the likelihood of stable and transient cavitation, the likelihood design performs best. The principal difference between the accuracy and the error metrics is that the first is based on binary classification between transient and stable cavitation, while the second allows real values that indicate the likelihood of cavitation type. From the training data presented in Section~\ref{sec: training data}, it was already clear that the differential equations and classifiers provide inconsistent results for a large proportion of the input parameters. Hence, the training data come with uncertainties in cavitation type, which are best handled with the likelihood design for machine learning.

Looking into the differences between the mean and voting techniques to achieve a final result in the ensemble and multi-objective designs, Table~\ref{tab:final_performance_metrics} shows that the mean methodology results in lower MAE and RMSE compared to the voting methodology. This implies that the mean method is more effective in situations with more intermediate cases, where for the same features or input parameters, the label or cavitation type is different across differential equations and classifiers.

Additionally, we notice that both the ensemble and multi-objective designs have an accuracy close to $82\%$. In contrast, the average accuracy presented in Figure~\ref{fig: accuracy for ensemble and multi-objetive} is $91\%$. This discrepancy arises because the accuracy of the combined designs is calculated in Table~\ref{tab:final_performance_metrics}, whereas the average accuracy across individual classifiers is considered in Figure~\ref{fig: accuracy for ensemble and multi-objetive}. Therefore, these are different calculations.

\subsubsection{Generalization}

So far, we presented performance metrics for machine learning predictions for input parameters within the range of the training data. However, we can also use the trained machine to predict the likelihood of cavitation for input parameters that are outside the training range. This generalization becomes increasingly challenging for parameters more distant from the training data. Here, we consider a generalization experiment where we leave out the $20\%$ of the highest input values for the initial radius. Hence, the training set consists of the lowest $80\%$ of values for the initial radius and the entire ranges for the other input parameters.

\begin{nolinenumbers}
\begin{table}[ht]
    \caption{Performance metrics of the machine learning designs on the test set for the generalization experiment, where the training set was selected as the $80\%$ of cases with the lower initial radius, while the test set comprised the $20\%$ of cases with the upper initial radius.}
    \label{table: generalization_results}
    \begin{tabular}{l c c c}
        \hline
        \textbf{Method} & \textbf{Accuracy} & \textbf{MAE} & \textbf{RMSE} \\
        \hline
        \text{Ensemble Mean} & 0.8295 & 0.0683 & 0.1007 \\
        \text{Ensemble Voting} & 0.8295 & 0.1705 & 0.4129 \\
        \text{Multi-objective Mean} & 0.8325 & 0.0683 & 0.1007 \\
        \text{Multi-objective Voting} & 0.8325 & 0.1675 & 0.4093 \\
        \text{Expansion Value} & 0.7036 & 0.2964 & 0.5444 \\
        \text{Likelihood Value} & 0.4585 & 0.0659 & 0.0941 \\
        \hline
    \end{tabular}
\end{table}
\end{nolinenumbers}

Table~\ref{table: generalization_results} presents the performance metrics for the generalization experiment. On first look, it is evident that the performance is similar to the cross-validation metrics presented in Table~\ref{tab:final_performance_metrics}. This confirms the effectiveness of generalization with machine learning. However, upon closer inspection, the accuracy is lower, and the error is higher in most cases. The reduced accuracy for generalization is expected behavior because we are testing machine learning predictions in cases unseen at the training phase. Yet, the performance deterioration is small. Hence, this experiment shows that machine learning provides reasonable estimates for cavitation experiments on input parameters just outside the training data.

\subsection{Cavitation likelihood charts}

As explained in the introduction, the exact physical properties of the bubbles and surrounding media are uncertain or unknown in many practical situations. Furthermore, acoustical parameters such as amplitude and frequency can often be chosen in laboratory experiments, and sonication protocols are optimized to achieve specific objectives for the onset and type of cavitation. Hence, cavitation threshold charts for a broad range of acoustical and physical parameters are needed to understand the likelihood of cavitation regimes in various situations. At the same time, trained machines quickly predict the variable of interest since they avoid solving the physical models for each set of input parameters. For this purpose, we trained our machine learning algorithms on the entire dataset and predicted the likelihood of transient cavitation for different input pairs of physical and acoustical input parameters. Specifically, we use the ensemble design and calculate the percentage of simulations (i.e., ODE and classifier combinations) that predict transient cavitation.

\begin{nolinenumbers}
\begin{figure}[ht]
    \begin{subfigure}[b]{0.4\textwidth}
        \includegraphics[width=\textwidth]{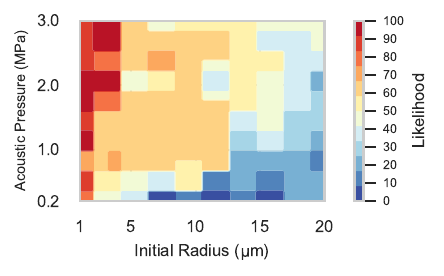} 
        \caption{1~MHz and 20 \textdegree C}
        \label{fig:subfig1}
    \end{subfigure}
    \begin{subfigure}[b]{0.4\textwidth}
        \includegraphics[width=\textwidth]{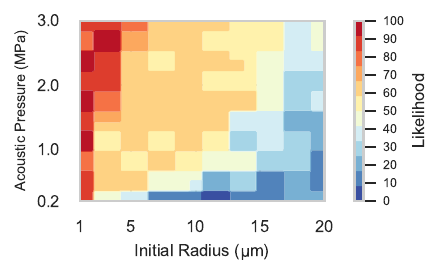} 
        \caption{1~MHz and 37 \textdegree C}
        \label{fig:subfig2}
    \end{subfigure}
    \begin{subfigure}[b]{0.4\textwidth}
        \includegraphics[width=\textwidth]{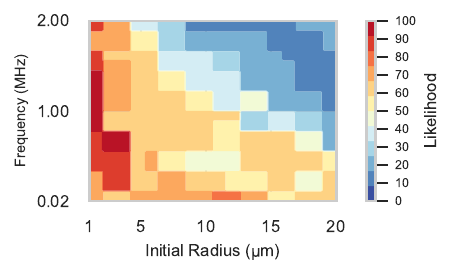} 
        \caption{1~MPa and 20 \textdegree C}
        \label{fig:subfig3}
    \end{subfigure}
    \begin{subfigure}[b]{0.4\textwidth}
        \includegraphics[width=\textwidth]{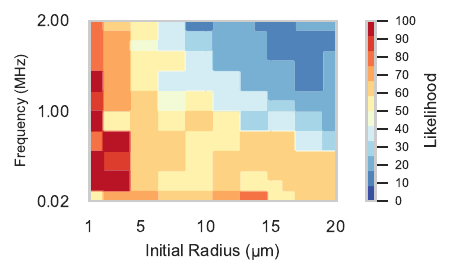} 
        \caption{1~MPa and 37 \textdegree C}
        \label{fig:subfig4}
    \end{subfigure}
    \caption{Cavitation charts in the form of heatmaps indicating the likelihood of transient cavitation as predicted by the machine learning algorithm with ensemble design. The independent variables are the temperature, initial radius, frequency, and acoustic pressure. The colors indicate the proportion, in percentage points, of transient cavitation predictions by the machine learning algorithm. The panels show different cases for two fixed input parameters.}
    \label{fig:heatmap_predictions}
\end{figure}
\end{nolinenumbers}

Figure~\ref{fig:heatmap_predictions} presents cavitation likelihood charts, where two input parameters are fixed, while the other two vary across their full range (see Table~\ref{tab: ranges for initial parameters for ML}). The heatmaps reveal the intricate, nonlinear effects of the bubble's equilibrium radius and acoustic parameters on bubble dynamics and cavitation regimes. The variability and block-like patterns in the heatmaps arise from the limited resolution of the training dataset and the inherently nonlinear nature of cavitation. 

According to the physical models of bubble dynamics~\cite{church2015theoretical,smirnov2021,deandrade2019,haqshenas2016modelling}, transient cavitation is more likely to occur at resonant bubbles (i.e., bubbles with resonance frequencies close to the driving frequency of the acoustic waves) and at higher acoustic pressures. For example, the resonance frequencies of bubbles with initial radii of \(1.7~\mu\)m and \(3.4~\mu\)m are approximately \(2~\text{MHz}\) and \(1~\text{MHz}\), respectively, in water at \(20~\text{°C}\). Thus, we expect a higher likelihood of transient cavitation for bubbles with the initial radii around a few microns at the $1-2$ MHz acoustic frequency range, as shown in Fig.~\ref{fig:heatmap_predictions} (a) and (c). Also, larger bubbles, with lower resonance frequencies, exhibit a small likelihood of transient cavitation at higher frequencies, as seen in Fig.~\ref{fig:heatmap_predictions} (c) and (d).  

At a fixed ultrasound frequency, the plots show a higher likelihood of transient cavitation as the pressure amplitude increases, especially for bubbles with resonance frequencies close to the ultrasound frequency, see Fig.~\ref{fig:heatmap_predictions} (a) and (b). These predictions are consistent with both experimental~\cite{neppiras1980,connolly1954,smirnov2021} data and theoretical findings~\cite{church2015theoretical,smirnov2021,deandrade2019,haqshenas2016modelling}.

The effect of temperature on bubble dynamics is complex. In general, as temperature increases, the surface tension, viscosity and liquid density decrease~\cite{deandrade2019,neppiras1980,smirnov2021}. This facilitates bubble oscillations and potentially increases the likelihood of cavitation. Similar observations can be made in Fig.~\ref{fig:heatmap_predictions} (b) and (d) compared to subfigures (a) and (c), where there are more blocks above $50\%$ likelihood at 37~\text{°C} compared to 20~\text{°C}. Despite nonlinear relationships between cavitation thresholds and the input parameters (i.e., acoustic parameters, the physical properties of the medium, and the initial bubble radii), the predictions from the developed machine learning model are consistent with the underlying physics of cavitation and in agreement (qualitatively) with experimental observations.

\section{Conclusions}
\label{sec: Conclusions}

In this study, we developed and evaluated machine learning algorithms for predicting cavitation regimes of air bubbles in liquids across a broad range of acoustic and material parameters. These algorithms were trained on simulated bubble dynamics data generated using various theoretical models, including the Rayleigh-Plesset, Keller-Miksis, and Gilmore differential equations. The trained models were tested across diverse scenarios with input parameters relevant to biomedical ultrasound and sonochemistry applications. Cross-validation on held-out test data achieved an accuracy of approximately 80\% in predicting the most likely cavitation regime.  

The best-performing machine learning model was used to compute cavitation threshold maps at different temperatures. The predictions integrate equally weighted contributions from multiple cavitation classifiers, including maximum bubble radius, maximum acoustic Mach number, Flynn’s criterion based on pressure and inertia functions, and the kurtosis of acoustic emissions. By incorporating data from multiple theoretical models and using multiple classifiers, the proposed approach provides a more comprehensive and statistically robust methodology compared to traditional cavitation threshold maps, which rely on a single physical model and threshold. The likelihood charts show good qualitative agreement with theoretical and experimental data published in the literature.  

The machine learning models developed in this study offer a fast, accurate, and reliable means of predicting cavitation likelihoods. However, the physical models used to generate training data assume radially symmetric oscillations of uncoated gas bubbles in viscous liquids. Future work aims to extend these models to more general scenarios, including: (i) coated bubbles oscillating in viscoelastic media, such as ultrasound contrast agents in soft tissue; (ii) multiple interacting bubbles; and (iii) scenarios requiring full simulations of the Navier-Stokes equations, such as asymmetric oscillations near boundaries, which are particularly relevant for microbubbles in blood vessels.  

\section*{Code and data availability}
The software code and data that support the findings of this study are openly available on GitHub (\url{https://github.com/trinidadgatica/Bubble-Cavitation-ML}).

\section*{Acknowledgment}
This work was financially supported by the Agencia Nacional de Investigación y Desarrollo, Chile [FONDECYT 1230642].

\section*{Author Declarations}
The authors have no conflicts to disclose.

\bibliography{aipsamp}

\end{document}